\documentclass[12pt,preprint]{aastex}

\usepackage{epsfig}

\begin{document}

\newcommand{\gsim}{\hbox{\rlap{$^>$}$_\sim$}} 
 
\title{On the X-ray lines in the afterglows of GRBs} 

\author{Shlomo Dado\altaffilmark{1}, Arnon Dar\altaffilmark{1} and
A. De R\'ujula\altaffilmark{2}}

\altaffiltext{1}{Physics Department and Space Research Institute,
Technion, Haifa 32000, Israel}
\altaffiltext{2}{Theory Division, CERN,CH-1211 Geneva 23, Switzerland}


\begin{abstract}
The observation of X-ray lines in the afterglow of GRB 011211 has been
reported, and challenged. The lines were interpreted as blue-shifted X-rays
characteristic of a set of photoionized ``metals'', located
in a section of a supernova shell illuminated by a GRB emitted a couple
of days after the supernova explosion. We show that the most prominent
reported lines coincide with the ones predicted in the ``cannonball''
model of GRBs. In this model, the putative signatures are Hydrogen lines,
boosted by the (highly relativistic) motion of the cannonballs (CBs). The
corresponding Doppler boost can be extracted from the fit to the observed
I-, R- and V-band light-curves of the optical afterglow of GRB 011211, so
that, since the redshift is also known, the line energies are --in the CB
model-- predicted. We also discuss other GRBs of known redshift which show
spectral features generally interpreted as Fe lines, or Fe recombination
edges. The ensemble of results is very encouraging from the CB-model's
point of view, but the data on each individual GRB are not good enough to
draw (any) objectively decisive conclusions. We outline a strategy for
X-ray observers to search for lines which, in the CB model, move
predictably from higher to lower energies.

\end{abstract}


\section*{Introduction} 

There is mounting evidence from late-time observations of the optical
afterglows (AGs) of relatively nearby (redshift $\rm z<1$) gamma ray bursts
(GRBs) that long duration GRBs are produced in the explosions of 
supernovae akin to
SN1998bw (Galama et al. 1998), by the ejection of ordinary baryonic
matter --essentially ionized Hydrogen-- in the form of plasmoids or 
``cannonballs'' (CBs), with very highly
relativistic Lorentz factors ($\gamma\sim 10^3$)
(Dar and De R\'ujula 2000a,b, Dado et al. 2002a,b,c), 
but otherwise
similar to the ones observed in quasars (Marscher et al. 2002) and
microquasars (e.g., Mirabel and Rodriguez 1994;  Belloni et al. 1997; 
Mirabel and Rodriguez 1999; Rodriguez and Mirabel 1999 and references
therein). The ejection of these cannonballs (CBs) close to the
line of sight makes their sky-projected motion appear extremely
superluminal\footnote{With the exception (Dar and De R\'ujula 2000a)
of the very close-by
GRB 980425, associated to SN1998bw, this ``hyperluminal'' motion is
not directly observable, but it was suggested (Dado et al. 2002d) that it 
gives 
rise to the observed scintillations in the GRB radio afterglows (Taylor et al.
1997) and may thereby be measurable.}. 

On Dec. 11, 19:09:21 UT 2001 the  long duration ($\sim 270$ s)
GRB 011211 was detected in the constellation Crater by BeppoSAX (Gandolfi
et al. 2001). Approximately ten hours after the GRB, its optical afterglow
was detected by Grav et al. (2001) and was followed by measurements of its
declining light-curve by Bloom et al. (2001), Jensen et al. (2001),
Holland et al. (2002), Bhargavi et al. 2001, Fiore et al. (2001), Burud et al. (2001) 
Covino et al. (2001); and Fruchter et al. (2001), who
also measured its redshift: $\rm z=2.141$, confirmed in turn by Gladders et al.
(2001).  The GRB's host galaxy was detected, with a red magnitude
$ \rm R_{host}=25.0\pm 0.3$, by Burud et al.  (2001). 

Observations with XMM-Newton of the X-ray afterglow of GRB 011211 started
at 06:16:56 UT on December 12, 2001, 11 h (39.6 ks) after burst, and lasted 27 ks 
(Reeves et al. 2002a).  The analysis of the X-ray spectrum revealed significant
evidence for emission lines only in the first 10 ks of observations. The
emission lines that were fitted to the first 5 ks data had energies of
$\rm 1.40 \pm 0.05\,keV$, $\rm 2.19 \pm 0.04\,keV$, $\rm 2.81 \pm
0.04\,keV$, $\rm 3.79 \pm 0.07\,keV$, and $\rm 4.51 \pm 0.12\,keV$,
in the GRB rest frame. They were interpreted by the observers as $\rm
K\alpha$ lines  from MgXI, SiXIV, SXVI, ArXVIII and CaXX,
blueshifted by  the motion at $\rm \beta= v/c= 0.086\pm 0.04$
of a shell ejected by a massive GRB progenitor in a supernova (SN) explosion 
having occurred a couple of days prior to the GRB.
In this interpretation, a section of the SN shell near the line of
sight was illuminated and reheated to a temperature of $\rm T \sim 4.5\pm
0.5 $ keV by the beamed GRB, and it emitted the blueshifted X-ray lines.

Borodzin and Trudolyubov (2002) have criticized the above interpretation
by noticing that the data showing the lines were accumulated during
the first 5 ks of observations, while the source was located near the
edge of a CCD chip, and that the lines disappeared as the satellite was 
subsequently
repositioned. Moreover, the background data collected over the
edge of the CCD show a very significant peak at the position of
the most prominent alleged line. Rutledge and Sako (2002) have also
criticized the significance of these data on statistical grounds.
Reeves et al. (2002b) have responded to these critiques and
insisted on the significance of their results, though they find
that a fit without the two lines of minimum and maximum energy
(attributed to MgXI and CaXX) is as good, or even a little better,
than the fit with all five lines.

We cannot enter into the above controversy. In what follows,
we discuss the data in Reeves et al. (2002a) at face value for, even if
their significance is weakened, they constitute a good stage
within which to discuss the predictions of the CB model 
(Dar and De R\'ujula 2000a) concerning
X-ray lines in GRB afterglows. We concentrate on GRB 011211 because
it has, so far, the best measured X-ray spectrum, but we also discuss
other GRBs of known redshift in which $\rm Fe$ lines and/or a
recombination edge have been claimed to be observed (GRB 970508: Piro et
al. 1998;  GRB 970828: Yoshida et al. 1999; 2001 and GRB 991216: Piro et
al. 2000).
In all cases these ``lines'' or ``recombination edges'' are not truly 
observationally established; we shall often refer to them as
spectral {\it features,} for the sake of precision.

The subject of X-ray lines in GRB
afterglows has attracted considerable theoretical attention:
Bottcher (2000),
Bottcher and Fryer (2001),  
Ghisellini et al. (2002),  
Kumar and Narajan (2002),
Lazzati et al.  (1999, 2002),
Meszaros and Rees (2001),  
Rees and Meszaros (2000), 
Vietri et al. (2001), 
Wang et al. (2002),
Weth et al. (2000). 

The interpretation of the X-ray features in Reeves et al. (2002a) 
as metal lines is not without
problems. First, the non-detection of the Fe $\rm K\alpha$ line was
argued to be due to the relatively long time it takes the $\beta$ decay
chain $\rm Ni^{56}\rightarrow Co^{56}\rightarrow Fe^{56}$ to produce $\rm
Fe^{56}$ in supernova explosions. This appears to be inconsistent with
the fact that the only X-ray lines with large flux and equivalent width
previously claimed to be detected in GRB afterglows were attributed to Fe lines: 
the BeppoSAX results for GRB 970508 (Piro et al. 1998) and GRB 000214
(Antonelli et al. 2000), the ASCA observations for GRB 970828 (Yoshida et al. 
1999; 2001) and the Chandra data on GRB 991216 (Piro et al. 2000).
Second, the fitted blueshift of the X-ray lines is supposedly due to the
beaming of the GRB radiation that illuminates only a small section of the
expanding SN shell, near the line of sight. The energy deposition time in 
such a segment is very short {\it in the
observer frame}: the GRB ejecta is moving initially with a large Lorentz
factor $\gamma$, and it overtakes a SN shell with an
estimated radius $\rm R_{SNS}\sim 10^{15}$ cm in $\rm t\sim (1+z)\,
R_{SNS}/c\, \gamma^2\leq 10$ sec of observer's time, for $\gamma\! >\! 100$.
The radiative cooling time of an
optically thin  SN shell with an electron density $\rm n_e \sim 10^{15}\, 
cm^{-3}$ and a temperature of 4.5 keV
is also extremely short: $\tau\ll 1$ s. The arrival times 
(in the observer frame) of recombination photons from the SN shell sector  
illuminated by a GRB jet of opening angle $\theta=20^o$ are spread over
 $\rm t=R_{SNS}\, (1+z)\, (1-cos\theta)/c=1.75$ h after the GRB, while the
putative XMM lines were observed 11 h after burst. These, and other puzzling
geometrical and physical details of the model, leave ample room for other
interpretations of the observations, should they be real.

An alternative interpretation (Dar and De R\'ujula
2001a) is that the spectral features observed in the
X-ray afterglows of GRBs are optical Hydrogen-recombination lines from 
the CBs that produce GRBs, Doppler shifted to the X-ray 
band by the CBs' highly relativistic motion ($\gamma\sim 10^3$) and
observed\footnote{Doppler-shifted Lyman, Balmer and
HeI lines have been detected from the mildly relativistic jets of
SS433 (e.g. Margon 1984; Kotani et al. 1996; Eikenberry et al. 2001).}
at very small angles ($\theta\sim 1/\gamma$). 

In this paper we show that the energies of the X-ray emission lines perhaps
detected in the afterglow of GRB 011211 happen to coincide with the 
energies of Hydrogen's Balmer and Lyman lines, 
redshifted by $\rm 1+z=3.141$ due to the cosmic expansion, and
blueshifted by a Doppler factor $\delta\sim 835$. This value of $\delta$
is, as we shall see, that of the CBs at the time of the observation of
the putative X-ray lines (some 11h after burst) 
obtained from a Cannonball-Model fit to the light-curve of the optical
afterglow of GRB 011211, prior to the reported X-ray observations. What
this means is that the  positions of the X-ray lines ---{\it predicted} in the
CB model--- coincide with the observed positions of the most
prominent spectral features in the data.

In discussing the previous indications for ``Fe'' lines in the AGs of 
GRBs 970508, 970828 and 991216
(Dar and De R\'ujula 2001a) we argued that they were
compatible with $\rm Ly\alpha$ emission, though the data were
insufficient to make a decisive distinction between a highly boosted 
hydrogen line and a merely redshifted Fe line. Here, we
rediscuss this issue in more detail, now that we also
have independent a-priori determinations of the values of 
$\delta$ in each individual GRB at the time of the corresponding X-ray 
observations. As for the case
of GRB 011211, all prominent features in the spectra coincide in
energy with lines that can be expected in the CB model. 

The CB model predicts that the X-ray lines should be relatively 
narrow  and move in
time from higher to lower frequencies, as the CBs decelerate
while ploughing though the interstellar medium (ISM).  The blending of
the emissions from unresolved CBs with somewhat different Doppler factors, 
and/or a poor energy resolution, may broaden the lines considerably and 
conceal the time-dependence of their energy. In the current data, the 
limited energy resolution and the required integration
over relatively long time-intervals would certainly have precluded
the observation of the predicted line motion.

\section*{CB model fit to the AG of GRB 011211} 
\noindent 
We do not give here a detailed description of the CB model, which we have
discussed {\it at nauseam} elsewhere (Dar and De R\'ujula 2000a,b, Dado et al.
2002a,d). We simply reproduce the formulae required for the 
analysis at hand.

In the CB model the afterglow has three origins: the ejected CBs, the
concomitant SN explosion, and the host galaxy (HG). These components are
usually unresolved in the measured GRB afterglow, so that the
corresponding light curves and spectra are measures of the cumulative energy 
flux density:  
\begin{equation} 
\rm F_{AG}=F_{CBs}+F_{SN}+F_{HG}\, .
\label{sum} 
\end{equation} 
The contribution from the host galaxy dominates
the light-curve at late times and was fitted (in each band) to the late
afterglow. The contribution of the supernova was modelled by assuming an
SN1998bw-like contribution placed at the GRB redshift, $\rm z=2.141$.
In this particular GRB, as in all the ones with redshift $\rm z>1.2$, the
SN contribution is too dim to be observable (Dado et al. 2002a).

In the CB model the jetted cannonballs are made of ordinary matter, mainly
hydrogen. For the first $\sim 10^3$ seconds of observer's time, a CB is still
cooling fast and emitting via thermal bremsstrahlung (Dado et al.  2002a), 
but after
that its emissivity is dominated by synchrotron emission from ISM 
electrons that penetrate in it. Integrated over frequency, this synchrotron 
emissivity is approximately equal to the energy deposition rate of the 
ISM electrons in
the CB\footnote{The kinetic energy of a CB is mainly lost to the ISM
protons it scatters; only a fraction $\rm\leq m_e/m_p$ is re-emitted by
electrons, as the AG.}. The electrons from the ISM that enter the CBs are Fermi 
accelerated there to a broken power-law energy distribution with a ``break'' 
energy (or more appropriately a ``bend'' energy) equal to their incident energy  
in the CBs' rest frame. In that frame, the electrons' synchrotron  emission 
(prior to attenuation corrections)
has an approximate spectral energy density  (Dado et al. 2002d):
\begin{equation}
\rm  F_{_{CB}}[\nu,t] = E_\gamma\,{dn_\gamma\over d\,E_\gamma}\sim f_0\;
{ (p-2)\, \gamma^2\, \over (p-1)\,\nu_b}{ [\nu /\nu_b]^{-1/2}
\over \sqrt{1+[\nu/ \nu_b]^{(p-1)}}}
\label{sync}
\end{equation}
where $\rm p\approx 2.2$ is the spectral index of the Fermi accelerated
electrons prior to the inclusion of radiation losses, $\rm f_0$ is an
explicit normalization constant proportional to the ISM baryon density
$\rm n_p$,
$\rm \gamma(t)=1/\sqrt{1-\beta^2}$ (with $\rm \beta=v/c$) is
the Lorentz factor of the CBs, and  
$\rm \nu_b \simeq 1.87\times 10^{3}\,[\gamma(t)]^3\,
[n_p/10^{-3}cm^{-3}]^{1/2}$ Hz
is the ``injection bend'' frequency in the CB rest frame\footnote{This bend 
frequency does  not correspond to the conventional
synchrotron ``cooling break''. It is produced by an {\it injection bend} 
in the high energy electron spectrum in the CB at the energy $\rm E_b = 
\rm\gamma(t)\ m_e\, c^2$ with which the ISM electrons enter the CB
at a particular time in its decelerated motion (Dado et al. 2002d).}.
The X-ray frequency domain in Eq.~(\ref{sync}) is always at
$\nu\gg\nu_b$, so that the expected spectrum is 
$\rm d\,n_\gamma/d\,E_\gamma
\approx E^{-\alpha}$, with a slope $\rm\alpha=(p+2)/2\simeq 2.1$.
The radiation emitted by a CB is Doppler-shifted   
and forward-collimated by its highly relativistic
motion, and redshifted by the cosmological expansion.
A distant observer sees a spectral energy flux:
\begin{equation}
\rm F_{obs}[\nu,t]\simeq
\rm { (1+z)\,\delta(t)^3\, R^2\, A(\nu,z)\,\over D_L^2}\,
F_{_{CB}}\left[{(1+z)\,\nu\over\delta(t)},{\delta(t)\,t\over 1+z}
\right]\, ,
\label{Fnuobser}
\end{equation}
where $\rm R$ is the radius of the CB (which
in the CB model tends to a calculable constant value 
${\rm R_{max}}={\cal{O}}(10^{14})$ cm, in minutes of observer's time),
$\rm A(\nu,z)$ is the total extinction along the line of sight to the GRB, 
$\rm D_L(z)$ is the luminosity
distance\footnote{The cosmological parameters we use are: $\rm H_0=65$
km/(s Mpc), ${\rm \Omega_M}=0.3$ and ${\rm \Omega_\Lambda}=0.7$.} and
$\rm\delta(t)$ is the Doppler factor of the light emitted by the CB:
\begin{equation}
\rm
\delta(t)=
{1\over\gamma(t)\,(1-\beta(t)\cos\theta)}
\simeq{2\, \gamma(t)\over 1+\theta^2\gamma(t)^2}\; ,
\label{doppler}
\end{equation}
where $\theta$ is the angle between the CB's direction of motion
and the line of sight to the observer.
The last approximation is valid in the domain 
of interest for GRBs: $\rm \gamma^2\gg 1$ and $\theta^2\ll 1$.
The total AG is the sum over CBs (or large individual GRB pulses) 
of the flux of Eq.~(\ref{Fnuobser}).

For an interstellar medium of constant baryon density $\rm n_p$, the
deceleration of the CBs results in a Lorentz factor, $\rm\gamma(t)$, that is 
given by (Dado et al. 2002a): 
\begin{eqnarray}
\rm \gamma&=&\rm\gamma(\gamma_0,\theta,x_\infty;t)
=\rm {1\over B} \,\left[\theta^2+C\,\theta^4+{1\over C}\right]\; ,\nonumber\\
\rm C&\equiv&\rm
\left[{2\over B^2+2\,\theta^6+B\,\sqrt{B^2+4\,\theta^6}}\right]^{1/3}\; ,
\nonumber\\
\rm B&\equiv&\rm
{1\over \gamma_0^3}+{3\,\theta^2\over\gamma_0}+
{6\,c\, t\over  (1+z)\, x_\infty}\; ,
\label{cubic}
\end{eqnarray}
where $\gamma_0=\gamma(0)$, and $\rm x_\infty=N_{_{CB}}/(\pi\, R_{max}^2\, n_p)$ 
characterizes the CB's slow-down in terms of
$\rm N_{_{CB}}$, its baryon number and $\rm R_{max}$,
its radius (it takes a distance $\rm x_\infty/\gamma_0$, typically of
${\cal{O}}(1)$ kp, for
the CB to slow down to half its original Lorentz factor).

\begin{figure}[]
\hskip 2truecm
\vspace*{0.3cm}
\hspace*{-1.6cm}
\plotone{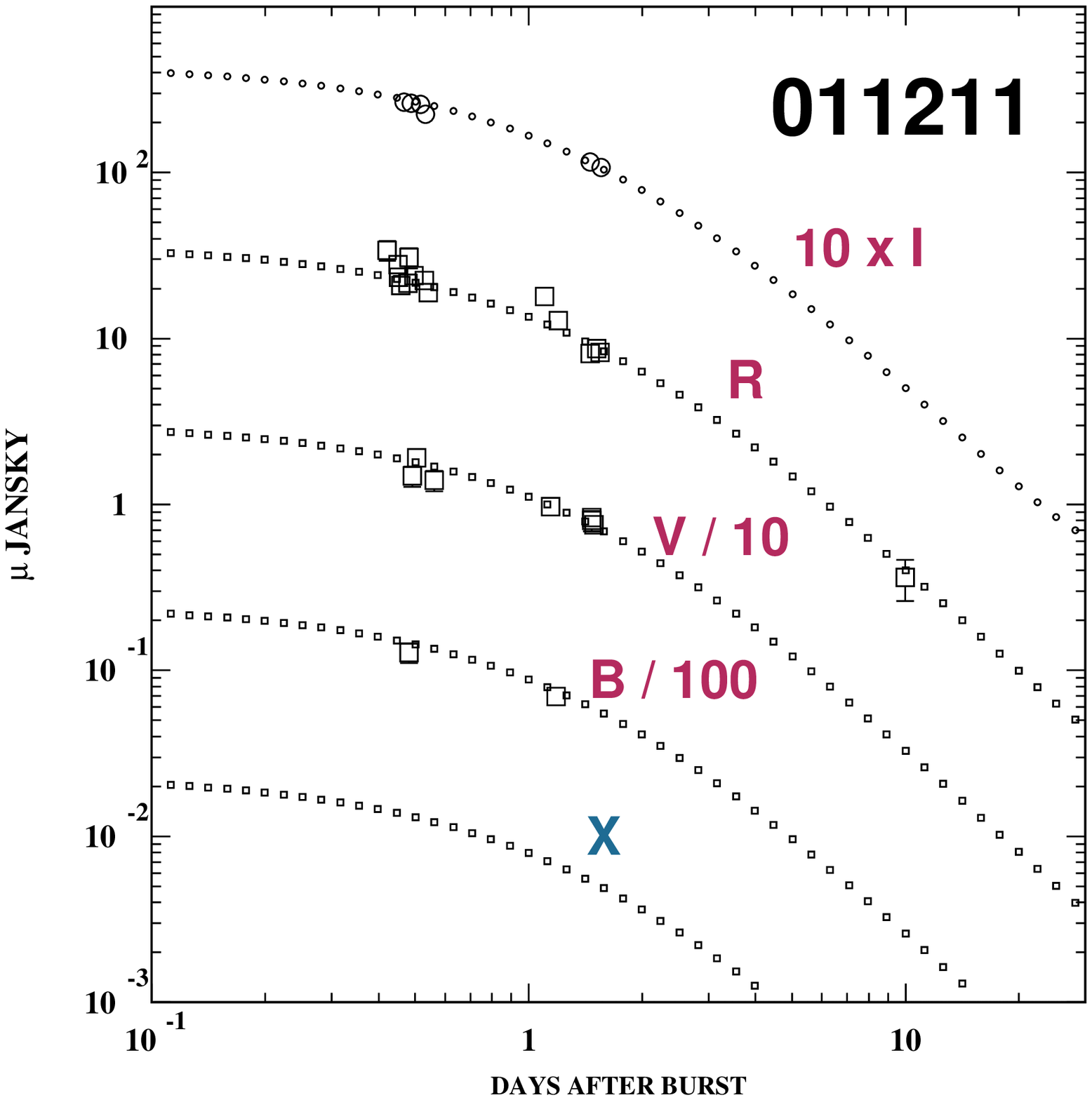}
\figcaption{Comparison between the observations in the I, R, V and
B bands of the optical afterglow of GRB 011211 and the CB model
fit as given by Eqs.~(\ref{sync}) to (\ref{cubic}).  The figure
shows (from top to bottom) 10 times the I-band, the R-band and 1/10
of the V-band and 1/100 of the B-band.  The line labelled X is the
predicted synchrotron contribution to the X-ray light curve. The
data are from Bhargavi et al. (2001) Burud et al.  (2001), Covino
et al. (2001), Fiore et al. (2001), Grav et al.  (2001), Jensen et
al.  (2001), Holland et al. (2002), and Fruchter et al.  (2001),
recalibrated with the observations of Henden et al.  (2002).
\label{ag011211}}
\end{figure}

In Fig.~(\ref{ag011211}), we show that the
optical afterglow  of GRB 011211 in the IRV bands is very well fitted in 
the cannonball model of GRBs, by use of Eqs.~(\ref{sync}) to (\ref{cubic}).
Besides the overall normalization, the  fit involves three parameters:
$\rm \theta=1.159\pm 0.005$ mrad; $\gamma_0=824\pm 2$, and 
$\rm x_\infty=0.271\pm 0.004$ Mpc. 
We have fixed the spectral index 
in Eq.~(\ref{sync}) to
 $\rm p=2.2$, since that value is compatible with the one fit,
in the same manner, to {\it all} of the optical, X-ray and radio AG light-curves 
of GRBs of known redshift (Dado et al. 2002a,d).  
There is a reason why in this fit we have not used the X-ray light curve
as part of the input. The contribution of the lines is a significant fraction
of the X-ray count rate, and they fade away rapidly in the observational
period between 0.46 and 0.77 days after burst. The lines are, in the CB model, a 
contribution that adds to the synchrotron-radiation described  in 
Eqs.~(\ref{sync}) to (\ref{cubic}). The X-ray light curve, thus, should
decline faster that its synchrotron component. The decline observed
in the measured interval is $\sim 50$\%, while the prediction is 32\% for
the synchrotron component, as determined from the fit to the optical
light curves.

It is consuetudinary in the GRB afterglow literature to quote values
of $\chi^2$ for the fits, and errors for the parameters. Yet, the
models are not  fundamental theories, but rough approximations
of no doubt hideously complicated phenomena.  Moreover, the dominant
contribution to the $\chi^2$ values very often originates in the spread
of almost simultaneous measurements; Fig.~(\ref{ag011211}) shows
this to be the case for GRB 011211, whose $\chi^2$/d.o.f. in our
CB-model fit is 1.4 for 26 data points\footnote{We have dealt with
the problem of partial data incompatibility with the method recommended
by the Particle Data Group (2000), tantamount in the cases at hand to
equating the difference between the extreme central values to a 
formal $\sim 2\,\sigma$ spread.}. In this fit, we approximate
the ISM-density and the CB radius by constants.  The tiny nominal
statistical errors of the fit, given in the previous paragraph, do
not include the systematic effects of deviations from these
approximations, nor do they reflect the systematic errors in the
data, involving the use of different detectors, assumptions about
absorption, etc.   Thus, even though our results here and elsewhere
are very good, it would be misleading to emphasize the ${\cal{O}}$
(1\%) high
``precision'' of our predicted line energies, since there is simply
no way to know what the ``true'' errors in the parameters are.
For other GRBs, we will not report the nominal parameter errors,
which are also minute.

\section*{The X-ray line data of GRB 011211}
\begin{figure}[]
\hskip 2truecm
\hspace*{-1.6cm}
\plotone{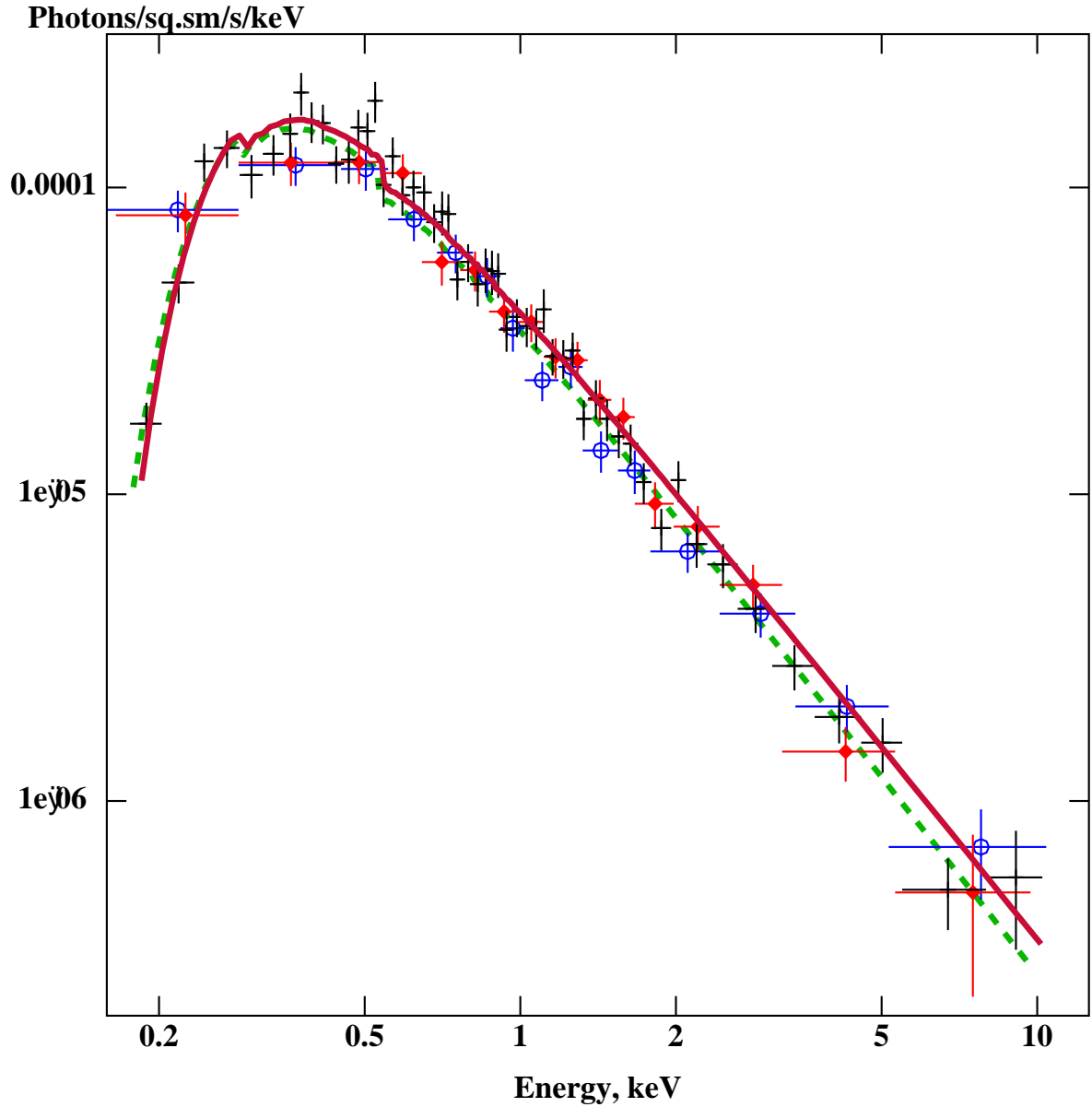}
\figcaption{X-ray spectrum of GRB 011211, as analized by Borodzin and 
Trudolyubov (2002). The data are from various detectors on board
XXM-Newton: MOS1 (circles), MOS2 (squares) and PN (no added symbols).
A power-law is only modified by absorption in the Galaxy. The dashed line 
is for a best-fit spectral index of 2.14, the continuous line is for the value
2.1 characteristic of the CB model.  
\label{211spectrum}}
\end{figure}

\noindent
The X-ray spectrum of this GRB has been well measured, in comparison
with previous cases. This is shown in Fig~(\ref{211spectrum}), which
we have borrowed from the data analysis by Borodzin and Trudolyubov
(2002). These authors find that the spectrum is compatible with
a power-law of slope $\alpha=2.14\pm 0.03$, modified only by absorption in the
Galaxy. A slope $\sim 2.1$ is expected in the CB model and 
---as extracted from the time-dependence of the optical and/or X-ray
AGs--- it is compatible with the observations for all GRBs of known
redshift (Dado et al. 2002a,b,c). In Fig~(\ref{211spectrum})
we also show how very compatible with the data at hand the expected
$\alpha \approx 2.1$ actually is.

\begin{figure}[]
\hskip 2truecm
\vspace*{.5cm}
\hspace*{-0.6cm}
\epsfig{file=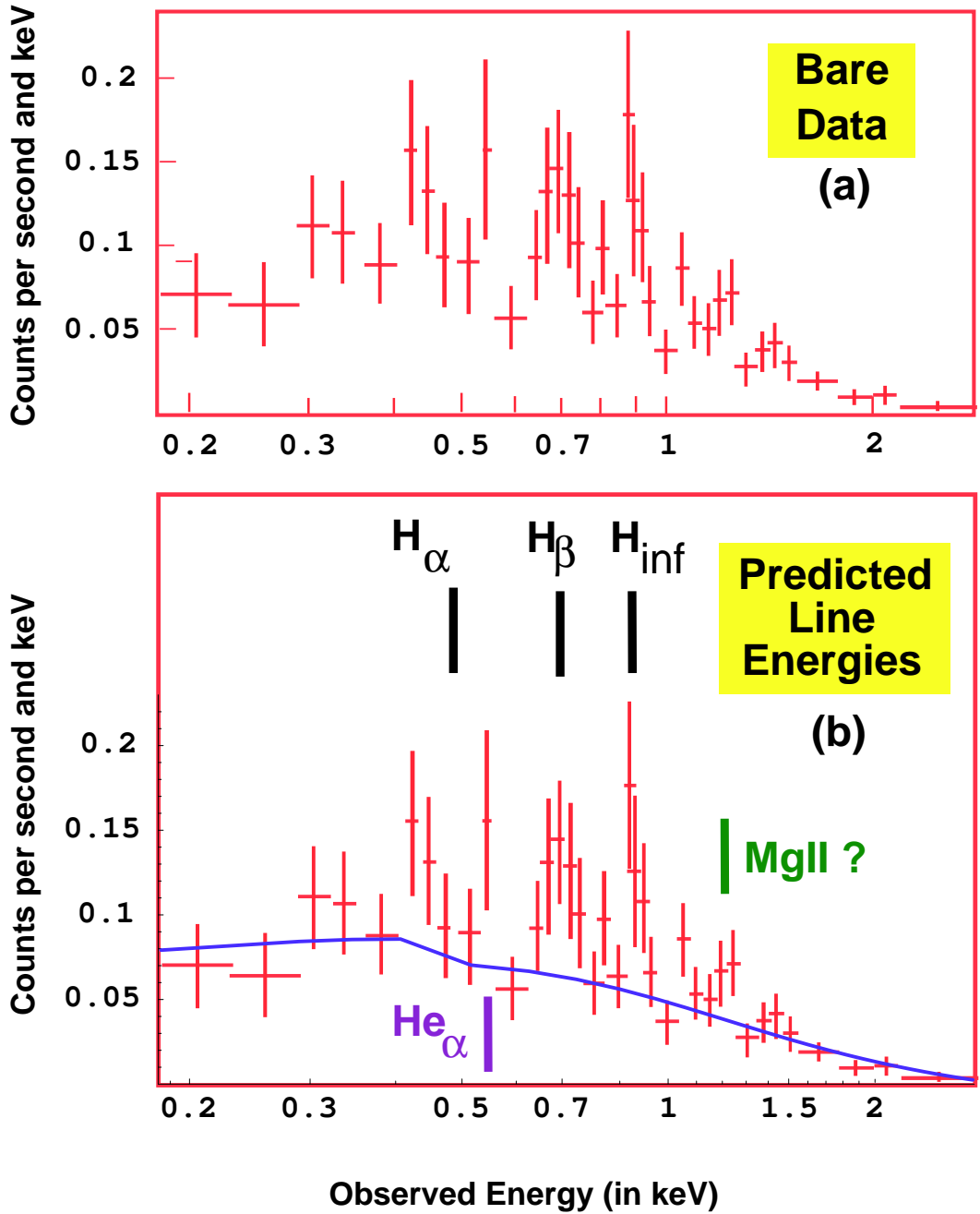,width=14cm}
\figcaption{(a) X-ray spectrum of GRB 011211 during the 5 ks of observations
in which putative line features were observed (Reeves et al. 2002).
(b) The same data with a ``background'' line scaled from a fit to the full 
27 ks of observations. The vertical lines are at the predicted positions
of lines in the CB model.
\label{211lines}}
\end{figure}

The observation and the properties of the lines reported in Reeves
et al.~(2002a) are model dependent; they are in particular very
sensitive to the continuum underlying the peaks in the data. This
can be seen in Fig.~(\ref{211lines}a), where we have redrawn the
data in Fig.~2 of Reeves et al. (2002a) without a model curve to
guide the eye (this figure reports data for the 5 ks interval in
which the lines were seen). Quite clearly, one can draw a smooth
continuum on this figure, above which the alleged lines would lose
much of their significance.  To draw such a continuum with as little
prejudice as possible we have first made a smooth fit to Fig.~1 of
Reeves at al.~(2002a), which displays the data on the complete 27
ks of observational time, where there are no significant line
features.   We have then redrawn
this continuum\footnote{The zig-zag feature around 0.5 keV, also
present in Fig. 1 of Reeves et al., (2002a) is presumably the effect
of the oxygen absorption edge in our Galaxy, somewhat smoothed by
resolution.} on top of the 5 ks data, with a normalization meant
to underemphasize the possible non-smooth deviations; the result
is shown in Fig.~(\ref{211lines}b).  
This procedure may not be a sophisticated data analysis,
but it is a sure way to account for detector-response and other
systematic effects, such as the energy-dependence of the detector's
effective area (all we are assuming is that these effects are the
same for the 5 ks and the 27 ks data sets).
Clearly, with the continuum ``background'' of Fig.~(\ref{211lines}b), 
the two alleged lines at $\sim1.21$ and $\sim1.44$
keV are not significant (these are the putative ArXVIII and CaXX
lines). Also, the other three smaller-energy lines, particularly
the lowest-energy one,
 are not very prominent\footnote{Recall that Reeves et al. (2002b)
also find that the inclusion of the lowest and highest energy lines
does not improve their fits.}.  The vertical lines in
Fig.~(\ref{211lines}b) are CB model expectations, which we proceed
to discuss.

\section*{Line emission in the CB model}
\noindent

As a CB ---in a time of ${\cal{O}}(1)$ s after it exits the transparent
outskirts of the shell of the SN associated with it--- becomes transparent
to the bulk of its enclosed radiation, its internal 
radiation pressure drops abruptly and its transverse expansion rate
is quenched by collisionless, magnetic-field-mediated 
interactions with the ISM (Dado et al. 2002a).
During this phase, the ISM electrons that enter the CB cool mainly by 
synchrotron emission. The synchrotron emission is 
partially reabsorbed by
the partially ionized CB through free-free transitions
at low radio frequencies and by bound--free and 
bound--bound transitions at optical frequencies (in the CB rest frame).
The CB plasma cools mainly by line emission from 
electron--proton recombinations. 
 
\subsection*{Line energies: the case of GRB 011211}
\noindent
At a given time $\rm t$, the CBs are viewed with a blue-shifting
Doppler factor $\rm \delta(t)$, so that a line of laboratory wavelength
$\rm \lambda_i$ would be observed at a redshift $\rm z$ to have an energy
uplifted by a ``boost'' factor $\rm B(t)$:
\begin{eqnarray}
\rm E_i(t)&=&\rm B(t) \, E_i^{lab}=B(t)\;{h\,c\over \lambda_i}\nonumber\\
\rm B(t)&\equiv&\rm{\delta(t)\over 1+z}\; ,
\label{Lineenergy}
\end{eqnarray}
with $\rm\delta(t)$ as in Eq.~(\ref{doppler}).
The parameters that we fit to the optical AG of GRB 011211, substituted
in Eqs.~(\ref{doppler},\ref{cubic}), result in $\rm \gamma(t)=668$ 
at $\rm t\sim 11$ h, the
time when the lines were seen. For a redshift $\rm z=2.141$, Eq.~(\ref{doppler}) 
implies $\rm \delta(t)\simeq 835$, so that lines at rest in the CBs would
be uplifted in energy by a factor $\rm B(t)\simeq 266$ at the time
of the observations.

\begin{figure}[]
\hskip 2truecm
\vspace*{.5cm}
\hspace*{-1.6cm}
\epsfig{file=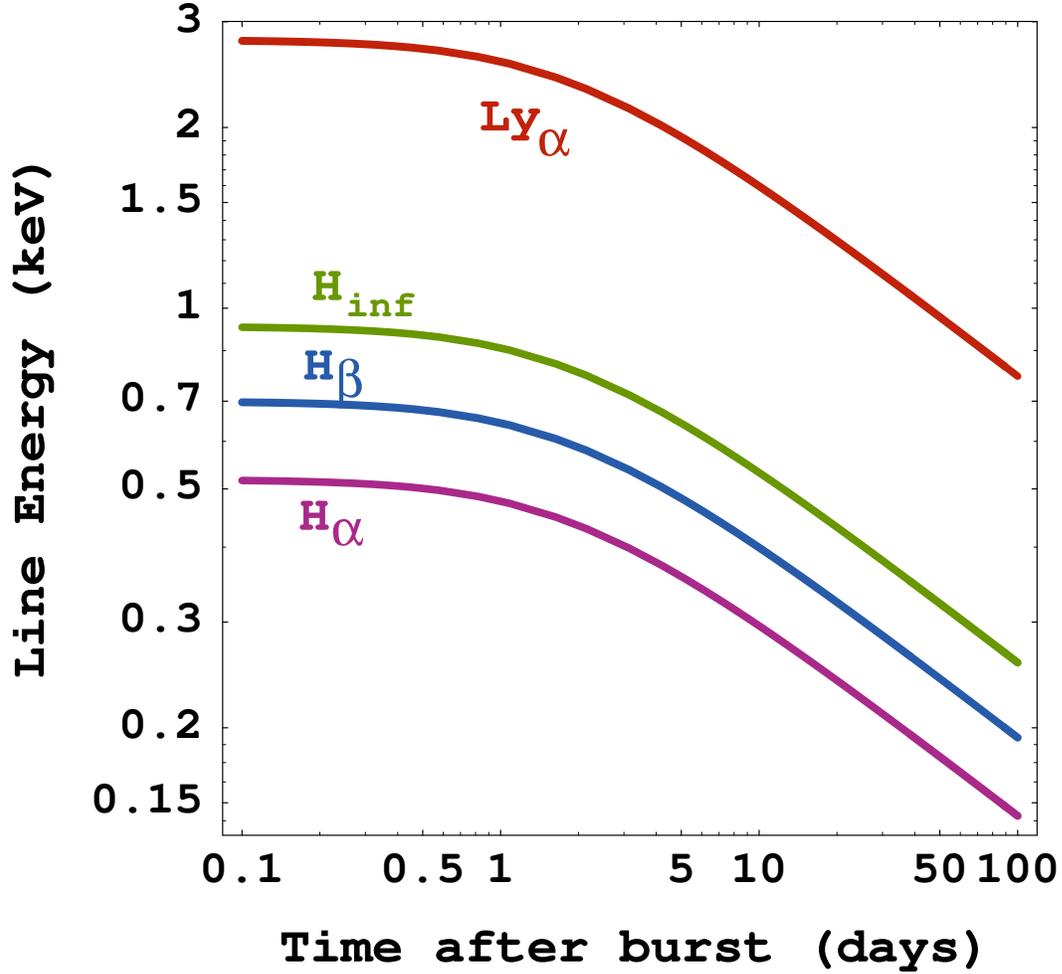,width=14cm}
\figcaption{Comparison between the predicted energies 
of the expected prominent lines in the X-ray afterglow
of GRB 011211 as function of time after burst,
as given by Eq.~(\ref{Lineenergy}), with the time-dependent Doppler
factor obtained from our CB-model fit to  its optical afterglow.
\label{linemigration}}
\end{figure}

In Fig.~(\ref{linemigration}) we show the predicted evolution of the
energy $\rm E_i(t)\propto\delta(t)$ of the $\rm H$ lines as a function of
time, for the case of GRB 011211. The X-ray observations of Reeves et al.
(2002a) lasted too little for the line motion to have an observable effect,
given their limited statistics and energy resolution. This is also the
case for all the other GRBs to be discussed below. 

In normal dense astrophysical plasmas, e.g., plasma clouds in the 
broad-line region of quasars (see, e.g., Laor et al. 1997
and references therein), the prominent Hydrogen lines are:
$\rm H\alpha [\lambda 6563]$, 
$\rm H\beta  [\lambda 4861]$, the higher energy Balmer lines accumulating
at $\rm H\infty [\lambda 3647]$, 
and the $\rm Ly\alpha[\lambda 1215.7]$ line.
The first three of these lines are, as one can see in Fig.~(\ref{211lines}b),
at the positions where there are, perhaps, indications in the data of an excess
over a smooth continuum: the predicted energies are 0.50, 0.68 and 0.91 keV,
while the fit of Reeves et al. (2002a) results in $0.45\pm 0.05$, $0.70\pm 0.02$
and $0.89\pm 0.01$ keV, respectively.
The $\rm Ly\alpha$ line ought to be uplifted in energy 
to $\sim 2.74$ keV, above the range shown in Fig.~(\ref{211lines}). Interestingly, 
there is a feature in the 27 ks data (Fig. 1 of Reeves et al. 2002a) which,
although it is also not very significant, sits at that very
point\footnote{We do not have access to
that figure in an e-friendly format, and it
is too complicated to reproduce by hand, or by scanning.}.
All these features have widths comparable with the experimental
resolution of somewhat less than 100 eV.

We have argued in Dar and De R\'ujula (2000, 2001a,b) and Dar et al.
(2000) that the ordinary matter
constituting cannonballs ought to be shattered by their violent
collision with the SN shell, and exit it in the form of unbound baryons
and electrons, so that the expected X-ray lines would be merely 
hydrogenic.  But it is quite possible that the collisions be somewhat
``cushioned'' (Hubbard and Ferry, in preparation, Hubbard 2002) such
as to leave some
nuclei unscathed, as the CBs gather SN shell material in their passage:
in their collisions with shell nuclei, the baryons or nuclei of the CBs
lose a considerable fraction of their initial Lorentz factor
(Dar and De R\'ujula 2001b), but they
may do it in many soft collisions, as opposed to a few hard ones.
It that case, one may expect to see also $\rm He$- or even
``metal'' lines, uplifted in energy by as much as the $\rm H$ lines are. 
The predicted position of the $\rm He\alpha[\lambda 5875]$
of $\rm HeI$
is also shown in Fig.~(\ref{211lines}b).
One quasi-degenerate
example of lines, prominent in the broad line region of quasars, is the pair
$\rm MgII[\lambda 2796.3;\, \lambda 2803.5]$, which in the case
at hand should appear at 1.19 keV, see Fig.~(\ref{211lines}b). 
This is  where Reeves et al. (2002a) 
claim to see Ar XVIII, at $1.21\pm 0.02$ keV. 
But the possible choices (other than for $\rm H$ and 
perhaps $\rm He$ and $\rm Ni$ lines) are far too vast  to
draw definite conclusions from these very scant data.
This is even more so in the ``standard''
interpretation, in which the overall line-shift is a free parameter.

\subsection*{Other GRBs with X-ray spectral ``features''}
\begin{figure}[]
\vspace*{.5cm}
\hspace*{1.6cm}
\epsfig{file=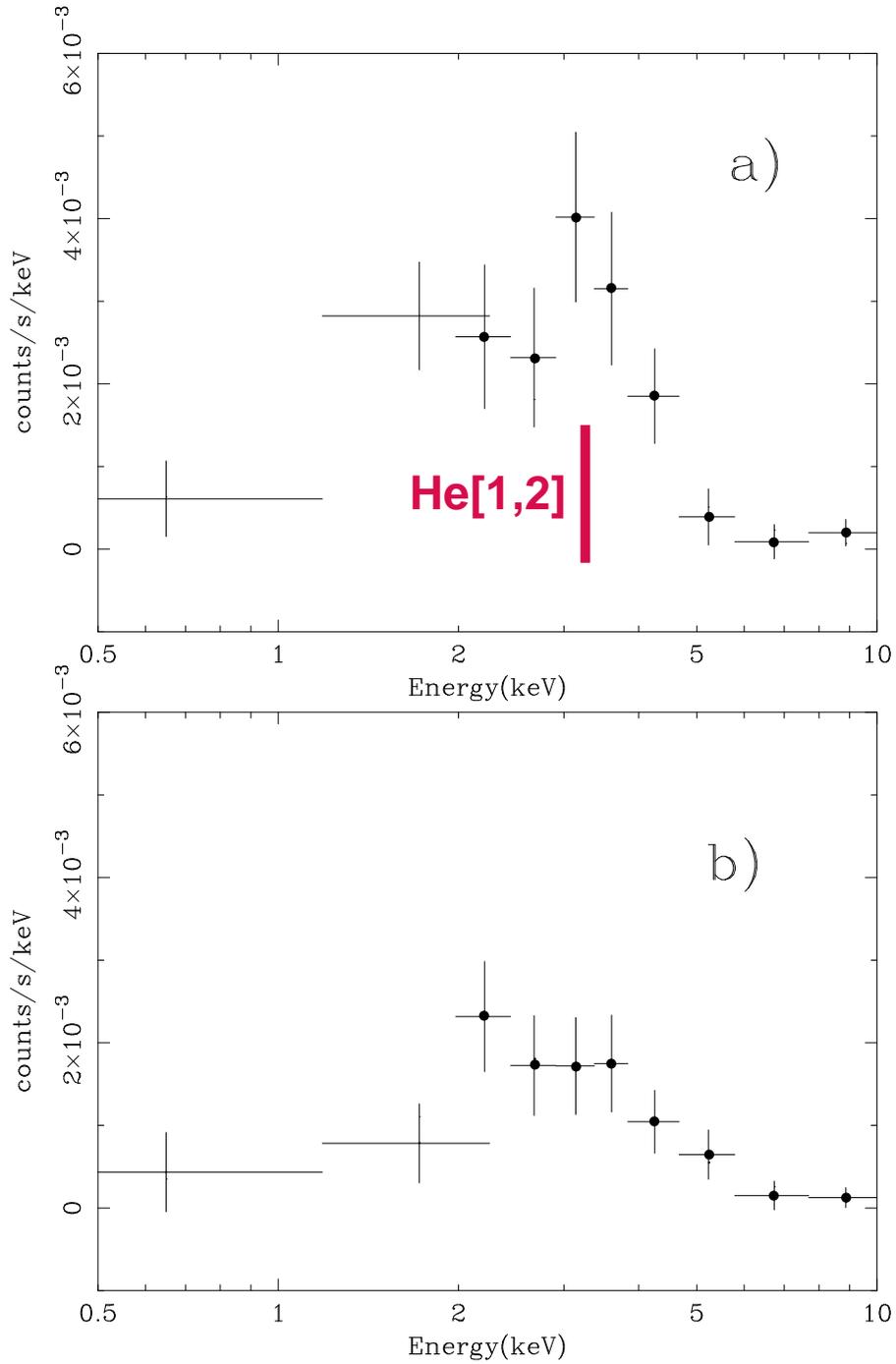,width=12cm}
\caption{The X-ray spectrum of GRB 970508. (a) In the first and (b)
in the second part of the observations (Piro et al. 1998). The vertical
line is at the position predicted in the CB model for the 
$\rm Ly\alpha$-like transition in  $\rm HeII$.
\label{508lines}}
\end{figure}

\begin{figure}[]
\hskip 2truecm
\hspace*{-0.5cm}
\plotone{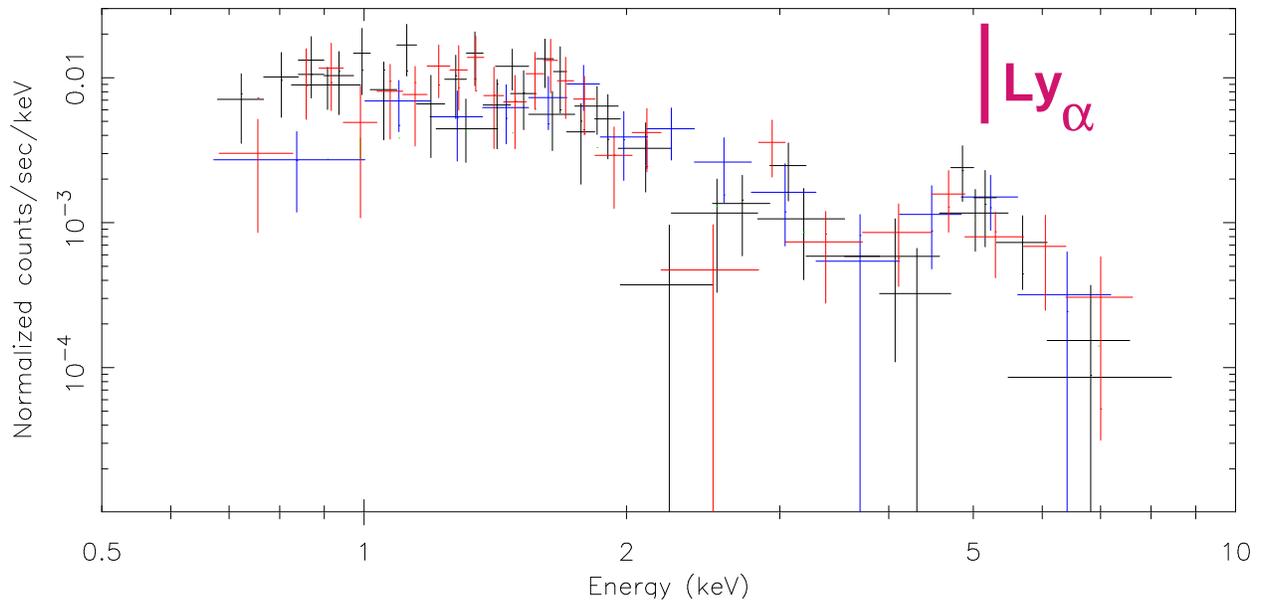}
\figcaption{The X-ray spectrum of GRB 970828 in the intermediate time-period
in which a putative line feature was observed (Yoshida et al. 2001).
The vertical line is at the position predicted in the CB-model for the
$\rm Ly\alpha$ transition.
\label{828lines}}
\end{figure}

\begin{figure}[]
\vskip 1cm
\hskip 2truecm
\hspace*{-0.5cm}
\plotone{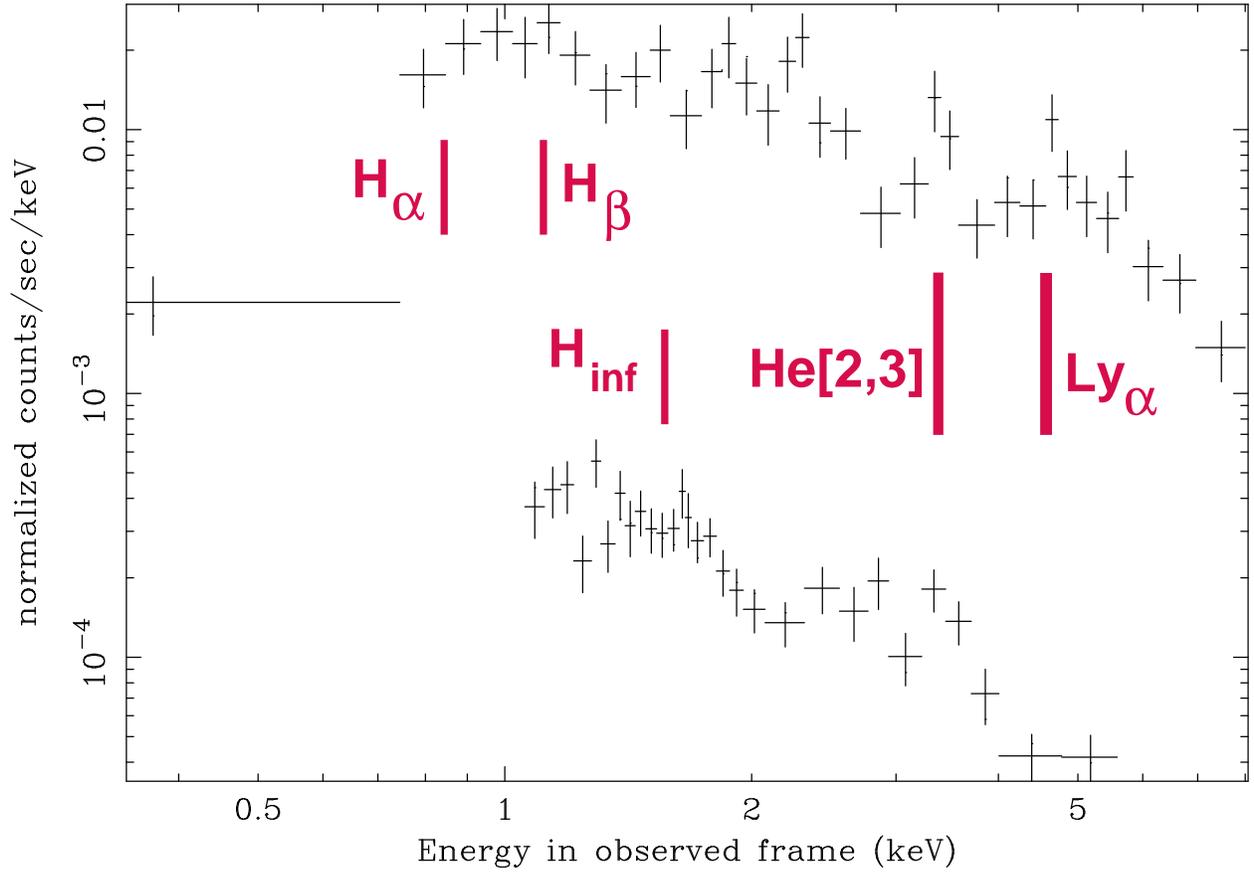}
\figcaption{The X-ray spectrum of GRB 991216. The top (bottom) spectrum is 
that of the ACIS-S (HETG) counter of
{\it Chandra} (Piro et al. 1999). The vertical lines are at the positions
predicted in the CB model (the $\rm He$ line is the $\rm H\alpha$-like
transition in  $\rm HeII$). 
\label{216lines}}
\end{figure}

\noindent
There are GRBs of known redshift in whose X-ray data the
observation of ``$\rm Fe$'' lines or recombination edges has
been claimed; in chronological order: GRB 970508 (Piro et al., 1999), GRB
970828 (Yoshida et al. 1999, 2001 and references therein) and GRB 
991216 (Ballantyne et al. 2002 and references therein). The corresponding
data are shown in Figs.~(\ref{508lines}, \ref{828lines}, \ref{216lines}),
from which we have, as for GRB 011211 in Fig.~(\ref{211lines}), eliminated 
theoretical lines that unavoidably ``guide the eye''. It is clear from these 
figures, without further ado, that
without a very good knowledge of the shape and magnitude of the
smooth continuum underlying the putative lines, it is not possible
to claim the observation of statistically convincing effects. 
The evidence for a line is more convincing in the case of
GRB 000214 (Antonelli et al. 2000), but its redshift is not known, precluding an 
explicit analysis.

The case for the observation of line features is presumably weakest
for GRB 970508, see Fig.~(\ref{508lines}). The upper panel is
the spectrum of the ``early'' X-ray AG, extending for some 30 ks after the 
start of the observations, at 6 hours after the GRB. The lower figure
shows the later data around 1 day after the burst. The feature at 
$\rm E\sim 3.4$ keV in the upper panel has been interpreted as
an $\rm Fe$ line at the GRB's redshift $\rm z=0.835$ (Piro et al.
2000). In Dado et al. (2002a) we have fit the 
optical AG of this GRB in the CB model, the resulting
parameters are $\theta=3.51$ mrad, $\gamma_0=1123$ and $\rm x_\infty=0.293$
Mpc\footnote{The value of $\rm x_\infty$ is for the early part of the AG, 
the time at which the X-ray 
line was possibly observed, which precedes the abrupt 
rise in this AG at $\rm\sim 1$ day, discussed in Dado et al. (2002a).}.
With these parameters, we obtain $\rm\delta(t_{obs})\simeq 142$
and $\rm B(t_{obs})=78$,
nearly constant through the X-ray observation time, as in 
Fig.~(\ref{linemigration}). Boosted by this $\rm B(t_{obs})$, one of the
potentially strong lines, the $\rm n=2$
to $\rm n=1$ transition in $\rm HeII$ would be at 3.17 keV,
where the feature is in the upper panel of Fig.~(\ref{508lines}). 
For this particular GRB, which is viewed at a relatively large angle,
$\delta$ and $\rm B$ are quite small, and other putative lines are at
sub-keV energies, where absorption appears to be very
strong. The data, however, are not good enough to extract conclusions from
the coincidence of the observed feature and the $\rm He$ line, nor from its 
$\rm Fe$-line interpretation.

GRB 970828, in spite of its being well localized (Remillard et al. 1997,
Smith et al. 1997, Marshall et al. 1997, Murakami et al. 1997, Greiner et al.
1997), had no detectable optical AG down to a magnitude $\rm R\simeq 23.8$
(Groot et al. 1998). Such ``orphan'' GRBs are expected in the CB
model, not only because of possible absorption, but because
the time at which the optical AGs begin to decline very fast is
extremely sensitive to the circumburst ISM density, 
and may be as short as ${\cal{O}}(10^{-2})$ days, see Fig.~(6) of Dado et
al. (2001). In that article, lacking optical data, we fit the X-ray light curve
of this GRB in the CB model, 
with the result that its parameters were $\gamma_0=1153$,
$\theta=0.86$ mrad and $\rm x_\infty=0.87$ Mpc
(these values are not as well determined as in GRBs with
well measured optical AGs). Yoshida et al. (1999, 2001)
analized the spectra of this GRB in three time intervals, in the
middle one of which, at $\rm t_{obs}\sim 1.2\times 10^5$ s, 
they found hints of structure, reproduced in Fig.~(\ref{828lines}).
These authors first ascribed the feature at $\rm E\sim 4.8$ keV
to the $\rm Fe$ $\rm K_\alpha$ line, which resulted in a prediction
of a redshift $\rm z=0.33$. The subsequently measured redshift
of the likely host galaxy is $\rm z=0.9578$ (Djorgosvski et al. 2001).
More recently Yoshida et al. (2001) attribute the feature to
a recombination edge of $\rm Fe$. The CB model fit results in 
$\rm \delta(t_{obs})\sim 1004$
 for which, at the GRBs redshift, $\rm B(t_{obs})\sim 513$, and 
a $\rm Ly\alpha$ line would be at a predicted $\rm E\sim 5.2$ keV which is,
as shown in Fig.~(\ref{828lines}), quite compatible with the position of
the apparent feature in the data.

In the case of GRB 991216, Piro et al. (2000) have interpreted the features
at $\sim 3.4$ keV and $>\! 5$ keV
of the X-ray spectra shown in Fig.~(\ref{216lines}) as the 6.7 keV
$\rm K_\alpha$ line of He-like $\rm Fe$, and a
$\rm Fe$ recombination edge, respectively. The reported significance
of the line is larger than $4\,\sigma$. On the other hand, Ballantyne et al. (2002) 
have analized in detail the line feature in a more specific
model. They report that the ``F-test'' significance of the $\rm K_\alpha$ line 
is 98\%, an explicit example of how difficult it is to convince oneself
that lines have actually been observed: a $\rm 2.33\,\sigma$
effect in a Gaussian distribution has the same significance\footnote{These
authors also find that an extra non-Galactic absorption ``is significant
{\it only} at the 97\% confidence level'' (the emphasis is ours, and 97\%
is $\rm 2.17\,\sigma$).}. For this GRB, the CB model fit to the optical AG 
results in $\gamma_0=906$,  $\theta=0.43$ mrad and $\rm x_\infty= 0.462$
Mpc, which imply, at the average observational time 
$\rm t_ {obs}\sim 39$ hours, $\rm\delta(t_ {obs})\sim 905$ 
and a boost $\rm B(t_{obs})\sim 448$ at the
GRB's redshift $\rm z=1.02$. For that predicted boost, 
there are no indications of deviations
from a smooth distribution at the positions of the $\rm H$ Balmer lines,
except, perhaps, for the corresponding recombination edge. But the $\rm n=3$
to $\rm n=2$ line of $\rm HeII$ and the $\rm H$ $\rm Ly\alpha$
line very snuggly coincide with the two allegedly significant
 features of the X-ray spectrum,
as shown in Fig.~(\ref{216lines}) (the alleged Fe line centers at 3.4 keV, the
predicted He line is at 3.39 keV). But, 
once again, the data are not precise enough to extract decisive inferences.

\section*{A rough way to search for lines}

The procedure we have discussed to predict
the Doppler factor, $\rm \delta(t)$, 
is elaborate: it involves fitting the available data with use of
Eqs.~(\ref{sum}) to (\ref{cubic}). There is an approximate and much 
simpler procedure that observers looking at a particular X-ray 
AG may find useful.

The observed X-ray frequencies are, in the
CBs' rest system,  always above the injection bend:
$\rm \nu \! \gg \! \nu_b$ in Eq.~(\ref{sync}). This implies that the 
observed energy flux of Eq.~(\ref{Fnuobser}) is:
\begin{equation}
\rm F_{obs}\propto \gamma^2\,\delta^3\,
\left[\nu \over \delta\right]^{-p/2}\, \nu_b^{(p-2)/2}
\propto \gamma^{3p/2-1}\,\delta^{3+p/2}\;\nu^{-p/2}
\label{approxflux}
\end{equation}
The index $\rm p$ can be fit to the X-ray spectrum of the AG
being studied. If this X-ray AG, as it usually the case, is observed
late enough for its light-curve to be an approximate
power law $\rm F_{obs}\propto t^{-\alpha}$,  $\alpha$ can 
be fit to the data. For this approximation to be good, 
it is necessary that 
$\rm \delta \approx 2\gamma$, that is $\theta^2\gamma^2\! \ll \! 1$
in Eq.~(\ref{doppler}), implying that Eq.~(\ref{approxflux})
further simplifies to $\rm F_{obs}\propto \delta^{2p+2}$. 
The conclusion is that $\rm \delta(t) \propto t^{-b}$ with 
$\rm b=\alpha/[2p+2]$ (for the theoretical $\rm p=2.2$,
$\rm \delta\propto b^{-\alpha/6.4}$, so that $\delta$ evolves roughly
as the 6.4-th root of the X-ray AG energy flux). This result
is approximate also in that Eq.~(\ref{approxflux}) describes
the synchrotron-radiated X-ray background, not the extra line
contribution. But if the latter is quite significant, the lines are
easier to find!

Knowing the time dependence of $\rm\delta(t)$ allows the observer
to ``stack'' the data taken at different times, in a search for lines
with an energy evolving with time as in Eq.(\ref{Lineenergy}), that is
$\rm E_i(t)\propto \delta(t)$. The  trick is to construct a 
time-integrated spectrum:
\begin{eqnarray}
\rm
\widetilde F(\widetilde \nu) & = & \rm
\sum_{t_{obs}}\;F_{obs}(\widetilde \nu,t_{obs})\, ,\nonumber\\
\rm
\widetilde \nu 
& \equiv & \rm \nu_{obs}\;\left[{t_{obs}\over t_0}\right]^b\; ,
\label{trick}
\end{eqnarray}
where $\rm t_0$ is an arbitrary reference time, e.g. the onset of the
observations: $\rm t_0=min[t_{obs}]$. In $\rm \widetilde F$ the CB-model's
lines occur at approximately fixed, time-independent scaled frequencies
$\widetilde \nu$, so that the lines ``stack up'' ---rather than drifting--- with time.

\subsection*{Line intensities}
\noindent
A detailed modelling of the line intensities is a very involved
problem. Here we can only offer a qualitative, order-of-magnitude
discussion of the subject.   

The recombination rate in a hydrogenic CB is (Osterbrock 1989):
\begin{equation}
\rm R_{rec}\simeq 6.0\times 10^{44}\, x^2\,
                  \left[{N_{_{CB}}\over 6\times 10^{50}}\right]\,
                  \left[{n_b\over 10^7\, cm^{-3}}\right]\,
                  \left[{T\over 10^4\,K}\right]^{-0.7}\, s^{-1}\,,
\label{Recombination}
\end{equation}
where $\rm N_{_{CB}}$ is the total baryon number of the CB,
$\rm n_b$ is its baryon density, and $\rm x$ is the fraction
of ionized hydrogen in the CB. 
The line emission luminosity in the CB rest frame is
$\rm R_{rec}\times 13.6\, eV\, .$ 
The corresponding radiation is boosted and relativistically beamed to an
observed energy flux: 
\begin{eqnarray}
\rm F_{lines} &\simeq & \rm 2.5 \times 10^{-11}\;(1+z)\; erg\, cm^{-2}\, s^{-1}
\nonumber \\ &\times&\rm n_{_{CB}}\,
x^2\,
                  \left[{D_L\over 2\times 10^{28}\, cm }\right]^{-2}
                  \left[{N_{_{CB}}\over 6\times 10^{50}}\right]
                  \left[{n_b\over 10^7\, cm^{-3}}\right]
                  \left[{T\over 10^4\,K}\right]^{-0.7}
                  \left[{\delta\over 10^3}\right]^{4},
\label{Recombinationflux}
\end{eqnarray}
where $\rm n_{_{CB}}$ is the number of CBs (or prominent GRB pulses),
and the luminosity distance  $\rm D_L$ is normalized
to its reference value
for $\rm z=1$ and our chosen cosmological parameters.

For the reference parameters to which we have normalized
Eq.~(\ref{Recombinationflux}), the flux
is comparable to those reported 
for the X-ray line-emissions in GRB afterglows. Its exact value 
depends rather weakly on temperature and quite
strongly on the ionization fraction  $\rm x$ in the CBs,
whose qualitative evolution can be assessed as follows. 
The bound-free cross section for photoionization of atomic hydrogen in its
$\rm n$-th excited state by photons with frequency above the ionization
threshold, $\rm \nu_n=3.29\times 10^{15}/n^2\, Hz$, is given by
$\rm \sigma_\nu(n)=n\,\sigma_1\,\bar{g}_n (\nu/\nu_n)^{-3}$, with
$\rm \sigma_1= {64\, \alpha\, \pi\, a_0^2/(3\, \sqrt{3}})$ $\simeq
7.91\times 10^{-18}$ cm$^2$ ($\rm a_0=0.53\times 10^{-8}$ cm
is the Bohr radius and $\rm \bar{g}_n$ is the Gaunt factor
for photo-absorption by hydrogen). Thus, a partially ionized CB 
with a typical radius $\rm R_{_{CB}}\simeq 2.5\times 10^{14}\, cm,$  and 
density $\rm n_b\sim 10^7\, cm^{-3},$ is opaque to optical
radiation. The recombination 
photons are repeatedly reabsorbed and reemitted  while diffusing out of 
the CB. The optical radiation of a CB (X-rays in the observer's  frame) is 
the sum of the line emission and the power-law synchrotron radiation from its  
surface. In a quasi-equilibrium 
state the ionization fraction is such as to keep the local recombination 
rate equal to the joint ionization rate by the synchrotron  
and recombination radiations\footnote{ 
The cooling rate of electrons via bremsstrahlung,
$\rm L_{brem}\simeq 1.43\times 10^{-27}\, \bar n_e\, T^{1/2}\, 
erg\, s^{-1}, $ is more than three orders of magnitude smaller
than the electron cooling rate via recombination and line emission.}.
The temperature is controlled by the same equilibrium, by
the CBs' surface energy loss and by the energy input from the continuing
collision of the CB and the ISM. It is difficult to ascertain without
a complete modelling of the problem. In Eq.~(\ref{Recombinationflux})
we have used a reference $\rm T$ so that the
maximum of the thermal distribution (at $\rm\sim\! 3\,T$) is
of the order of magnitude of the line energies of Hydrogen,
whose transitions dominate the thermal energy transport within the CB.

Initially, the ionization is close to maximal and the line radiation of
Eq.~(\ref{Recombinationflux}) results in a flux comparable to that of the power-law-behaved synchrotron radiation in the X-ray band. 
Later, when $\rm \gamma(t)$
decreases, equilibrium between the ionization and recombination rates 
 results in a rapid  decline of line emission:
the recombination rate is $\rm\!\propto\!x^2$ and the photo-ionization
rate is $\rm\propto\! (1-x)\,\gamma^2$, so that when $\rm x$ is small,
$\rm x\!\propto\!\gamma$, and the recombination rate decreases like 
$\gamma^2$. Moreover, the diffusion
time of recombination photons becomes very long as $\rm x\to 0$
which results in strong suppression of line emission.
The derivation of an exact X-ray spectrum and its time dependence would
require very complicated radiation-transport calculations which are beyond
the scope of this paper.

\section*{Conclusions}
\noindent
We have studied an alternative (Dar and De R\'ujula 2001a)
to the interpretation by Reeves et al. (2002a)
of the X-ray data of XMM-Newton on GRB 011211. Unlike the quoted authors,
we have not for the moment studied in detail the very involved question of
the predicted absolute and relative intensities of the lines, which is very
model dependent (the density profile, ionization level and temperature
of a CB, as well as their time dependence, are quite complicated issues).
But we have shown that, in the CB model, the positions of the lines
are predictable and happen to coincide with the meager
evidence for most of them. 
In the CB model
long-duration GRBs are associated with SNe that are compatible with
an approximately SN1998bw-like standard candle. Reeves
et al. (2002a) adduce that their data
supports this association; we contend that ---if it does--- it is not for the
reasons they advance, but because their observations are consistent
with CB-model expectations.

We have shown that the Fe-line candidates observed in three other
GRBs could very well be $\rm H$ or $\rm He$ lines, again predictably
boosted by the very fast motion of cannonballs. The individual data
on each of the four GRBs that we have discussed is inconclusive, but the
overall consistency of the CB-model interpretation of their X-ray
spectral features is encouraging. In the CB model the presence of X-ray
lines which ---case by case--- have predictable energies,
is a very natural possibility. In contrast, in the
other scenarios that have been discussed, the X-ray lines require 
in every instance the introduction of ad-hoc and sometimes rather exotic
hypothesis on the surroundings of the GRB engine.

With better data it ought to be possible to distinguish the lines expected
in the CB model from the ones of the standard GRB paradigm(s). Not only
the line positions can, in the CB model, be foretold; but also their widths
should be narrow, and predictably time dependent (Dar end De R\'ujula 2001a).

\acknowledgements{We thank Ehud Behar and Ari Laor for useful
comments. The support of the Asher Fund for Space Research at the Technion
is gratefully acknowledged.}

\newpage

\end{document}